\magnification=\magstep1
\tolerance=500
\rightline{ 14 May, 2010}
\bigskip
\centerline{\bf Hamiltonian Map to Conformal Modification of}
\centerline{\bf Spacetime Metric: Kaluza-Klein and TeVeS}
\bigskip 
\centerline{ Lawrence Horwitz,$^{1,2,3}$Avi Gershon$^1$,  and Marcelo
  Schiffer$^2$} 
\smallskip
\centerline{$^1$ School of Physics, Tel Aviv University, Ramat Aviv
  69978, Israel} 
\centerline{$^2$ Department of Physics, Ariel University Center of
  Samaria, Ariel 40700,Israel}
\centerline{$^3$Department of Physics, Bar Ilan University, Ramat Gan
  52900, Israel}
\bigskip
\noindent{\it Abstract}
\par It has been shown that the orbits of motion for a wide class of
non-relativistic Hamiltonian systems can be described as geodesic
flows on a manifold and an associated dual by means of a conformal
map
.  This method can be
applied to a four dimensional manifold of orbits in spacetime
associated with a relativistic system. We show that a relativistic
Hamiltonian which generates Einstein geodesics, with the addition of a
world scalar field, can be put into correspondence in this way with another
Hamiltonian with conformally modified metric.  Such a construction
could account for part of the requirements of Bekenstein for achieving
the MOND theory of Milgrom in the post-Newtonian limit. The
constraints on the MOND theory imposed by the galactic rotation
curves, through this correspondence, would then imply constraints on the
structure of the world scalar field. We then use the fact that a  Hamiltonian 
with vector gauge fields results, through such a conformal map, in a 
Kaluza-Klein type theory, and indicate how the {\it TeVeS} structure
of Bekenstein and Saunders can be 
put into this framework. We exhibit a class of
infinitesimal gauge transformations on the gauge fields ${\cal
U}_\mu(x)$ which preserve the Bekenstein-Sanders condition 
${\cal U}_\mu{\cal U}^\mu =
-1$. The underlying quantum structure giving rise to these gauge
fields is a Hilbert bundle, and the gauge transformations induce a
non-commutative behavior to the fields, {\it i.e.} they become of
Yang-Mills type. Working in the infinitesimal gauge neighborhood of the initial
Abelian theory we show that in the Abelian 
limit the Yang-Mills field equations provide residual nonlinear
terms which may avoid  the caustic singularity found by Contaldi {\it
et al}.
 
\bigskip
\noindent PACS 04.50.Cd,04.50.Kd,98.62.-g,95.35.+d,95.30.Sf
\bigskip
 \noindent{\bf 1. Introduction}
\smallskip
\par The Hamiltonian [1]
$$ K = {1 \over 2m} g^{\mu\nu} p_\mu p_\nu, \eqno(1)$$
with Hamilton equations (written in terms of derivatives with respect
 to an invariant world time $\tau$ [2])
$$ {\dot x}^\mu = {\partial K \over \partial p_\mu} = {1 \over m}
g^{\mu\nu} p_\nu \eqno(2)$$
and
$$ {\dot p}_\mu = -{\partial K \over \partial x^\mu} = -{1 \over
  2m}{\partial g^{\lambda\gamma} \over \partial x^\mu} p_\lambda p_\gamma
 \eqno(3)$$
lead to the geodesic equantion
$$ {\ddot x}^\rho = -{\Gamma^\rho}_{\mu\nu} {\dot x}^\nu {\dot x}^\mu,
\eqno(4)$$
where what has appeared as a compatible connection form
${\Gamma_\rho}^{\mu\nu}$ is given by
$$ {\Gamma^\rho}_{\mu\nu}= { 1\over 2} g^{\rho \lambda} \bigl( {\partial
  g_{\lambda\mu} \over \partial x^\nu} + {\partial
  g_{\lambda\nu} \over \partial x^\mu} - {\partial
  g_{\mu\nu} \over \partial x^\lambda}\bigr). \eqno(5)$$
\par These results are tensor relations over the usual diffeomorphisms
  admitted by the manifold $\{x^\mu\}$; writing the Hamiltonian in terms 
of $(2)$, we see that the
 invariant interval on an orbit is
  proportional, through
  the constant Hamiltonian, to the square of the world time of
  evolution on the orbit, {\it i.e.},
 $$ds^2 = {2\over m}K d\tau^2. \eqno(6)$$
\par We shall first study, in the following, a generalization of $(1)$
consisting of the addition of a scalar field $\Phi(x)$.  The presence
of such a scalar field may be considered as a gauge compensation field
for the $\tau$ derivative in the evolution term of the 
covariant generalization of $(1)$ in the St\"uckelberg-Schr\"odinger
equation [3], an energy distribution not directly associated with 
electromagnetic radiation in the usual sense.  We then follow the
method of ref.[4] to show that there
is a corresponding Hamiltonian ${\hat  K}$ with a conformally modified
metric, and no explicit additive
scalar field, which has the form of the construction of Bekenstein and
Milgrom [5] for the
realization of Milgrom's MOND program (modified Newtonian dynamics) [6]
for achieving the observed galactic rotation curves. This simple
form of Bekenstein's theory (called RAQUAL), which we discuss in some detail
below for the sake of simplicity and clarity in the development of the
mathematical method, does not properly account for causality and
gravitational lensing; the theory has been further
developed to include vector fields (which we shall call
Bekenstein-Sanders fields) as well({\it TeVeS}) [7], which has 
been relatively successful in accounting for these problems.
It has been shown[8], moreover,  that a gauge type Hamiltonian, with 
Minkowski metric and both vector and scalar fields results, under a conformal
map, in an effective Kaluza-Klein theory. We shall
indicate here (using a general Einstein metric) how the {\it TeVeS}
 structure can 
emerge in terms of a Kaluza-Klein theory in this way. It is
essential in this construction that the Bekenstein-Sanders fields be
considered as gauge fields. As a realization of this possibility, we exhibit 
an infinitesimal gauge transformation on the underlying
quantum theory for which the vector fields, which we shall call ${\cal
U}^\mu(x)$, emerge as gauge compensation fields, such that, as
required by the {\it TeVeS} theory the property ${\cal U}^\mu {\cal U}_\mu
= -1$ [7] is preserved under such gauge transformations. The
corresponding quantum theory then has the form of a Hilbert
bundle and, in this framework, the gauge fields are of (generalized)
Yang-Mills type [9]. Working in the infinitesimal neighborhood of a gauge
in which the fields are Abelian, we show that in the limit the
contributions from the nonabelian sector provide nonlinear terms in
the field equations which may avoid the caustic singularity found by
Contaldi {\it et al} [10].  Further investigation of this structure  will
be given in a subsequent publication. 
\par  For both the RAQUAL and the $TeVeS$ theories, the
correspondence between $K$ and ${\hat K}$ implies a relation between
the conformal factor in ${\hat K}$ and the world scalar field $\Phi$,
and thus a possible connection between the so-called dark matter
problem and a dark energy distribution represented by $\Phi$.
\bigskip
\noindent{\bf 2. Addition of a scalar potential and conformal equivalence}
\smallskip
\par The addition of a scalar potential to the Hamiltonian $(1)$, in
the form
$$ K = {1 \over 2m} g^{\mu\nu} p_\mu p_\nu + \Phi(x), \eqno(7)$$
 leads,
according to the Hamilton equations, to the geodesic
 equation\footnote{$^1$}{Note that $(8)$ does not admit an equivalence
 principle, but $(11)$, arising from $(9)$ does.} 
$$ {\ddot x}^\rho = -{\Gamma^\rho}_{\mu\nu} {\dot x}^\nu {\dot x}^\mu
-{1 \over m} g_{\rho \nu}{\partial \Phi \over \partial x^\nu}.
\eqno(8)$$
\par Now, consider the Hamiltonian (we carry out the calculations explicitly
here since we shall have need of some of the intermediate results)
$$ {\hat K} = {1\over 2m} {\hat g}^{\mu\nu}(y) p_\mu p_\nu. \eqno(9)$$
It follows from the Hamilton equations that
$$ {\dot y}^\mu = {\partial {\hat K}\over \partial p_\mu} = {1 \over
m} {\hat g}^{\mu\nu} p_\nu ,$$
so that
$$p_\nu = m{\hat g}_{\mu\nu} {\dot y}^\mu \eqno(10)$$
and 
$$ {\dot p}_\mu= - {\partial {\hat K}\over \partial y^\mu}
= -{1 \over 2m}{\partial {\hat g}^{\lambda \gamma} \over \partial
y^\mu} p_\lambda p_\gamma .$$
As in $(4)$, it then follows that
$${\ddot y}^\mu = - {\hat \Gamma}^\mu_{\lambda \sigma} {\dot
y}^\lambda {\dot y}^\sigma, \eqno(11)$$
where, as for $(4)$, 
$${\hat \Gamma}^\mu_{\lambda \sigma} = {1 \over 2}{\hat g}^{\mu\nu}
\bigl\{ {\partial  {\hat g}_{\nu \sigma} \over \partial y^\lambda} + 
{\partial  {\hat g}_{\nu \lambda} \over \partial y^\sigma}-{\partial 
 {\hat g}_{\lambda \sigma} \over \partial y^\nu}\bigr\}. \eqno(12)$$

\par We now establish an equivalence between the Hamiltonians $(7)$
and $(9)$  by assuming the momenta
$p_\mu$ equal at every moment $\tau$ in the two descriptions. With the 
constraint
$$  {\hat K} = K = k, \eqno(13)$$ 
if we assume the conformal form
$$ {\hat g}^{\nu \sigma}(y) = \phi(y) g^{\nu \sigma}(x), \eqno(14)$$
it follows that
$$\phi(y)(k-\Phi(x)) = k. \eqno(15)$$
\par The relation $(15)$ is not sufficient to construct $y$ as a function
of $x$, but if we impose the relation
$$ \delta x^\mu = \phi^{-1}(y) \delta y^\mu \eqno(16)$$
between variations generated on position in the two coordinate
systems, it is sufficient to evaluate derivatives of $\phi(y)$ in
terms of derivatives with respect to $x$ of the scalar field
$\Phi(x)$[11](see also [12]).
We review this construction briefly below. 
\par  We remark that the construction based on Eqs. $(9)$ and
$(14)$ admits the same family of diffeomorphisms as that of $(7)$,
since $\phi$ is scalar. Under these 
diffeomorphisms, both  $g_{\mu\nu}$ and ${\hat g}_{\mu \nu}$  
are second rank tensors, and by construction of the connection forms,
$(4)$ and $(11)$ are covariant relations.
\par To see how these derivatives are constructed on the constraint
hypersurface determined by $(15)$, let us, for
brevity,  define
$$F(x) \equiv {k \over k-\Phi(x)}, \eqno(17)$$
so that  the constraint relation $(15)$ reads
$$ \phi(y) = F(x). \eqno(18)$$
Then, since variations in $x$ and $y$ are related by $(16)$,
$$ \phi(y+\delta y) = F(x +\delta x) \cong F(x) + \delta x^\mu
{\partial F(x) \over \partial x^\mu}.\eqno(19)$$
To first order in Taylor's series on the left, we obtain the relation
$${\partial \phi(y) \over \partial y^\mu} = \phi^{-1}(y) {\partial F(x)
\over \partial x^\mu}. \eqno(20)$$
 We may therefore define a
derivative, restricted to the constraint hypersurface
$${{\tilde \partial} F(x) \over {\tilde \partial} y^\mu} = \phi^{-1}(y)
{\partial F(x) \over \partial x^\mu} \eqno(21)$$
The Leibniz relation follows easily for the product of functions,
{it e.g.}, for $\phi(y) g^{\mu\nu}(x)$.
\par In a similar way, the second derivative can be obtained from $(19)$
by recognizing that the variation in $x$ is to be computed at the
point $y + \delta y$. Keeping terms of second order in the expansion
of both sides, one can define the second derivative restricted to the
constraint hypersurface defined by $(18)$; although it appears that a
second derivative defined by $(21)$ would not be symmetric, both the
derivative of $(21)$ and the second derivative computed from $(19)$ on
the constraint hyperfurface agree and are symmetric [11], i.e.,
$${ {\tilde \partial}^2 F(x) \over {\tilde \partial} y^\mu {\tilde
\partial} y^\nu}={ {\tilde \partial}^2 F(x) \over {\tilde \partial}
y^\nu {\tilde\partial} y^\mu} \eqno(22)$$
This implies that the restricted derivative defined by $(21)$ behaves
as a {\it bona fide} derivative on the constraint hypersurface,
admitting the consistent coexistence of the coordinates $x$ and $y$
related by $(15)$. We will not have further use of $(22)$ here, primarily
relevant for the calculation of stability criteria through geodesic
deviation (the second derivative occurs in the curvature tensor).
\par  In the following, we complete our argument of equivalence
by reconstructing the equations of motion following from the Hamilton
equations applied to $(7)$, {\it  i.e.}, Eq. $(8)$.
\par We begin our reconstruction, in analogy with the procedure used in
the nonrelativistic problem[8], by defining the new variable $z_\mu$ such
that
$$ {\dot z}_\mu = {\hat g}_{\mu\nu}(y)  {\dot y}^\nu. \eqno(23)$$
Substituting $ {\dot y}^\nu = {\hat g}^{\mu\nu}(y){\dot z}_\mu$ into  
 $(11)$, the $\tau$ derivatives of ${\hat g}^{\mu\nu}(y)$ generate 
terms that cancel two of the  terms in ${\hat \Gamma}^\mu_{\lambda
\sigma}$, leaving
$$ {\ddot z}_\nu = {1 \over 2} {\partial {\hat g}_{\lambda \sigma}
\over \partial y^\nu} {\dot y}^\lambda {\dot y}^\sigma. \eqno(24)$$
Now, substituting for ${\dot y}^\lambda$ from $(23)$, and using the identity
$$ {\hat g}^{\gamma\lambda}{ \partial {\hat g} _{\lambda \sigma}\over
\partial y^\nu} {\hat g}^{\sigma \rho}= -{ \partial {\hat g} ^{\gamma
\rho}\over\partial y^\nu}, \eqno(25)$$
we find
$$ {\ddot z}_\nu = - {1 \over 2}{ \partial {\hat g} ^{\gamma
\rho}\over\partial y^\nu} {\dot z}_\gamma {\dot z}_\rho. \eqno(26)$$  
Finally, from the variational type argument we used above,
$$ \eqalign{{\hat g}^{\rho \gamma}(y + \delta y) - {\hat g}^{\rho
 \gamma}(y)
 &= {\partial {\hat
g}^{\gamma \rho} \over \partial y^\nu} \delta y^\nu \cr
&=   {\partial {\hat g}^{\rho\gamma} \over \partial y^\nu} {\hat
g}^{\nu \lambda} \delta z_\lambda,\cr} \eqno(27)$$ 
so that 
$$ {\partial {\hat g}^{\rho\gamma} \over \partial y^\nu} {\hat
g}^{\nu \lambda} =  {\partial {\hat g}^{\rho\gamma} \over \partial
z_\lambda}$$
or
$$ {\partial {\hat g}^{\rho\gamma} \over \partial y^\nu}= {\hat
g}_{\nu\lambda} {\partial {\hat g}^{\rho\gamma} \over \partial
z_\lambda} \eqno(28)$$
We therefore have the alternative form
$$ {\ddot z}_\nu = - {1 \over 2} {\hat g}_{\nu\lambda}{\partial {\hat
g}^{\rho\gamma} \over \partial z_\lambda}  {\dot z}_\rho {\dot
z}_\gamma. \eqno(29)$$
This result constitutes a ``geometric'' embedding of the Hamiltonian
motion induced by $(7)$ in the same way as for the nonrelativistic
case. Substituting the explicit form of ${\hat g}^{\rho \gamma}$ in
terms of the original Einstein metric from $(14)$, one obtains
$$ {\ddot z}_\nu = - {1 \over 2}g_{\nu\lambda}{\partial g^{\rho\gamma}
\over \partial z_\lambda} {\dot z}_\rho {\dot z}_\gamma - {1 \over 2}
\phi^{-1} g_{\nu\lambda} {\partial \phi \over \partial z_\lambda}
g^{\rho\gamma}{\dot z}_\gamma {\dot z}_\rho \eqno(30)$$ 
The second term contains the potential field, as in the Hamilton
equations, but the first term does not contain the full connection form.
We may finally, however, define a ``decontraction'' of the connection
in $(30)$ using the Einstein metric. In fact, since according to
$(16)$, ${\dot y}^\nu = \phi {\dot x}^\nu$,  and by $(23)$,
$$ {\dot z}_\mu = {\hat g}_{\mu\nu} {\dot y}^\nu = \phi^{-1}
g_{\mu\nu} {\dot y}^\nu, \eqno(31)$$
it follows that
$$ {\dot z}_\mu = g_{\mu\nu} {\dot x}^\nu. \eqno(32)$$
Making this substitution in $(30)$ leads explicitly, taking into
account the $k$ shell constraint $(13)$ and the form of $(7)$,
to the equation $(8)$.  We have thus completed our demonstration of
the equivalence between the purely metric form of the Hamiltonian $(9)$
and the Hamilton $(7)$, for which the relation $(29)$
corresponds to a dynamics generated by a compatible connection form,
and constitute a ``geometric'' embedding of the original Hamiltonian motion.
\par  Our interest
in this section has been in relating  
the Hamiltonian $(7)$ to the simplest Bekenstein-Milgrom form of MOND,
without concern in the development of this simplified case for lensing or 
causal effects, for which a
$TeVeS$ type theory
would be required.  In the next Section, we indicate how a $TeVeS$
theory can be 
generated in this framework, {\it i.e.}, as a result of a conformal map.   
\bigskip
\noindent{\bf 3. {\it TeVeS} and Kaluza-Klein Theory}
\smallskip
\par In this section, we show that the {\it TeVeS} theory can be cast
into the form of a Kaluza-Klein construction. There has recently been a 
discussion [8], from the point of view of conformal correspondence, of
the equivalence of a relativistic Hamiltonian with an electromagnetic
type gauge invariant form [3] (here 
$\eta^{\mu\nu}$ is
the Minkowski metric $(-1,+1,+1,+1)$) 
$$ K= {1\over 2m} \eta^{\mu\nu} (p_\mu - ea_\mu)(p_\nu - ea_\nu)-ea_5,
\eqno(33)$$  
where the $\{a_\mu\}$, as fields, may depend on the affine parameter
$\tau$ as well as $x^\mu$, and the $a^5$ field is necessary for the
gauge invariance of the $\tau$ derivative in the quantum mechanical
Stueckelberg-Schr\"odinger equation, with a Kaluza-Klein theory.
As remarked in this work, Wesson [13]
and Liko [14] , as well as previous work on this structure[3],  have 
associated the source of the $a_5$ field with mass density.
 A Hamiltonian of the form
$$ {\hat K}= {1\over 2m} {\hat g}^{\mu\nu}(p_\mu - ea_\mu)(p_\nu -
ea_\nu)\eqno(34)$$ 
can be put into correspondence with $K$ by taking ${\hat g}^{\mu\nu}$
to be
$$ {\hat g}^{\mu\nu} =  \eta^{\mu\nu} {k \over k+ea_5}, \eqno(35)$$
where $k$ is the common (constant) value of $K$ and ${\hat K}$. In
this correspondence, the equations of notion generated by ${\hat K}$
through the Hamilton equations, 
have extra terms, beyond those provided by the connection form
associated with ${\hat g}^{\mu \nu}$, due to the presence of the gauge
fields. These additional terms can be identified as 
belonging to a
connection form associated with a five dimensional metric, that of a 
Kaluza-Klein theory.
 \par We may apply the same procedure to the Hamiltonian
$$ K= {1\over 2m} g^{\mu\nu} (p_\mu -\epsilon {\cal U}_\mu)(p_\nu -
\epsilon {\cal U}_\nu) + \Phi,
\eqno(36)$$ 
where $g^{\mu\nu}$ is an Einstein metric, $\Phi$ is a world
scalar field, and ${\cal U}_\mu$ are to be identified with the
Bekenstein-Sanders fields for which[6]  ${\cal U}_\nu{\cal U}^\nu =
-1$, with   ${\cal U}^\mu = g^{\mu\nu} {\cal U}_\nu.$  

  We discuss in Section 4 a class of gauge
transformations on the wave functions of the underlying quantum theory
for which the ${\cal U}_\mu$ arise as gauge compensation fields. 
\par Let us define, as in Eq. $(35)$, the conformally modified metric
$$\eqalign{ {\hat g}^{\mu\nu} &=  g^{\mu\nu} {k \over k-\Phi}\cr
                          &\equiv e^{-2\phi} g^{\mu\nu}. \cr}\eqno(37)$$
The ``equivalent'' Hamiltonian
$$ {\hat K}= {1\over 2m} {\hat g}^{\mu\nu} (p_\mu - \epsilon{\cal
U}_\mu)(p_\nu - \epsilon{\cal U}_\nu)
\eqno(38)$$
then generates, through the Hamilton equations, an equation of motion
which corresponds to the geodesic equation for an effective
Kaluza-Klein metric, as in ref.[8].    
\par  Now, consider the Hamiltonian 
 $$ K_K = { 1 \over 2m}{\tilde g}^{\mu\nu}p_\mu p_\nu, \eqno(39)$$ 
 with the Bekenstein-Sanders metric[7]
$$ {\tilde g}^{\mu\nu} = e^{-2\phi} (g^{\mu\nu} + {\cal U}^\mu {\cal
U}^\nu) - e^{2\phi}{\cal U}^\mu {\cal U}^\nu \eqno(40)$$
The Hamiltonian $K_K$ then has the form
$$ K_K = e^{-2\phi}g^{\mu\nu}p_\mu p_\nu - 2\sinh{2\phi}
({\cal U}^\mu p_\mu)^2, \eqno(41)$$

\par Let us now define a Kaluza-Klein type metric of the form obtained in 
[7], arising from the equations of motion generated by $(38)$, 
$$ g^{AB} = \left(\matrix{ {\hat g}^{\mu\nu} & {\cal U}^\nu \cr
                {\cal U}^\mu & g^{55}\cr }\right).\eqno(42)$$
Contraction to a bilinear form with  the (5D) vectors $p_A = \{p_\lambda,
 p_5\}$, with indices $\lambda = \nu$ on the right and $\lambda = \mu$
 on the left, one finds
$$ g^{AB} p_A p_B = {\hat g}^{\mu\nu} p_\mu p_\nu + 2 p_5 (p_\mu {\cal
U}^\mu) + (p_5)^2 g^{55}. \eqno(43)$$
 If we take 
$$ p_5 = -{(p_\mu {\cal U}^\mu)\over g^{55}}\bigl( 1
\pm\sqrt{1-2g^{55}\sinh 2\phi}\bigr), \eqno(44)$$
then the Kaluza-Klein theory coincides with $(41)$, {\it i.e.}, 
$$ K_K = {1 \over 2m}g^{AB} p_A p_B. \eqno(45)$$
 As discussed by Wesson [13], Kaluza [15] chose $g_{55}= {\rm const.}$
 for consistency with electromagnetism, while Wesson [13] makes the more
 general choice of a world scalar field.  In particular, the value
 $g^{55}=0$ is well defined (as in [8]). 
\par Since the fields ${\cal U}^\mu$ are
timelike unit vectors[7],
$(p^\mu {\cal U}_\mu)$ corresponds, in an
appropriate local frame, to the energy of the particle, close to its
mass in the case of a nonrelativistic particle, or to the frequency in
the case of on-shell photons.  It clearly remains to understand more
deeply the apparently {\it ad hoc} choice of $p^5$ in $(44)$ in terms of a $5D$
canonical dynamics, along with the structure of the $5D$ Einstein
equations for $g_{AB}$ that follow from the geometry associated with
$(45)$. We shall study these questions in a succeeding paper.
\bigskip
\noindent{\bf 4 The Bekenstein-Sanders Vector Field as a Gauge Field}
\smallskip
\par Essential features of the Bekenstein-Sanders field [7] of the {\it TeVeS}
theory are that it be a local field, {\it i.e.}, ${\cal U}_\mu(x)$,
and there is a normalization constraint
$${\cal U}^\mu {\cal U}_\mu = -1,\eqno(46)$$ 
so that the vector is timelike.  To preserve the normalization
condition $(46)$ under gauge
transformation, we shall study the construction of a class of gauge 
transformations which essentially moves the 
${\cal U}(x)$ field on a hyperbola with
 a Lorentz transformation (at the point $x$).
\par  If we think of our underlying
quantum structure, which generates the gauge field, as a fiber
bundle with base $x^\mu$, then we must think of the transformation
acting  in such a way that the absolute square (norm) of the wave
function attached to the base point $x^\mu$ preserves its
value [9].
\par An analogy can be drawn to the usual Yang-Mills gauge [9] on $SU(2)$,
where there is a two-valued index for the wave function
$\psi_\alpha(x)$. The gauge transformation in this case is a two by two matrix
function of $x$, and acts only on the indices $\alpha$. The condition
of invariant absolute square (probability) is 
$$ \sum_\alpha \vert\sum_\beta  U_{\alpha \beta} \psi_\beta \vert^2 = \sum
 \vert \psi_\alpha \vert^2 \eqno(47)$$
\par  Generalizing this structure, one can take the indices $\alpha$ 
to be continuous, so that $(47)$ becomes
$$ \int (d{\cal U}) \vert \int (d{\cal U}')U({\cal U},{\cal
U}')\psi({\cal U}',x)\vert^2 = 
\int (d{\cal U})  \vert \psi({\cal U},x)\vert^2, \eqno(48)$$
implying that $U({\cal U},{\cal U}')$ is a unitary operator on a Hilbert space
$L^2(d{\cal U})$. Since we are assuming that ${\cal U}_\mu$ lies on an orbit
determined by $(48)$, the measure is
$$ (d{\cal U}) = {d^3 {\cal U} \over {\cal U}^0},\eqno(49)$$
{\it i.e.}, a three dimensional Lorentz invariant integration measure.
  \par Moreover, the Lorentz transformation on ${\cal U}_\mu$ is
generated by a non-commutative operator, and therefore the gauge 
transformation is
non-Abelian. We demonstrate the resulting noncommutativity of the
operator valued fields, ${\cal U}'$, after an infinitesimal gauge
transformation of ths type, explicitly below.
\par  This construction is somewhat
similar to the treatment of the electromagnetic potential vector and
its time derivative as oscillator variables in the process of second
quantization of the radiation field (the energy density of the field
is given by these variables in the form of an oscillator).
\par We now examine the gauge condition:
$$ (p_\mu -\epsilon {\cal U}'_\mu) U\psi = U (p_\mu - \epsilon {\cal
U}_\mu)\psi \eqno(50)$$
Identifying $p_\mu$ with $-i \partial /\partial x^\mu$, and cancelling
the terms $Up_\mu \psi$ on both sides, we obtain
$$ {\cal U}'_\mu = U{\cal U}_\mu U^{-1} - {i \over \epsilon} {\partial U \over
  \partial x^\mu} U^{-1}, \eqno(51)$$
in the same form as the Yang-Mills theory [9].  It is evident in the
Yang-Mills theory, that due to the matrix nature of the second term, 
the field will be algebra-valued, resulting in the usual structure
of the Yang-Mills non-Abelian gauge theory. Here, if the
transformation $U$ is a Lorentz transformation, the numerical valued
field ${\cal U}_\mu$ would be carried, in the first term, to a new
value on a hyperbola. However, the second term may well be operator
valued on $L^2(d{\cal U})$, and thus, as in the Yang-Mills theory, 
${\cal U}'^\mu$
would become nonabelian.
 \par It follows from $(51)$ that the field strengths 
$$ f_{\mu\nu} =  {\partial {\cal U}_\mu \over \partial x^\nu} -
 {\partial {\cal U_\nu} \over \partial x^\mu} +i\epsilon[{\cal U}_\mu,
 {\cal U}_\nu]\eqno(52)$$  
 are related to the the field strengths in the transformed form
$$ f'_{\mu\nu} =  {\partial {\cal U}'_\mu \over \partial x^\nu} -
 {\partial {\cal U}'_\nu \over \partial x^\mu} +i\epsilon[{\cal U}'_\mu,
 {\cal U}'_\nu]
 \eqno(53)$$ 
according to
$$f'_{\mu\nu}(x) = Uf_{\mu\nu}(x)U^{-1}, \eqno(54)$$
just as in the finite dimensional Yang-Mills theories.
\par This result follows from writing out, from $(51)$,
$$\eqalign{{\partial {\cal U}'_\mu \over \partial x^\nu}&= {\partial U
\over \partial
  x^\nu} {\cal U}_\mu U^{-1} + U{\partial {\cal U}_\mu \over \partial
  x^\nu} U^{-1}
+ U{\cal U}_\mu {\partial U^{-1} \over \partial x^\nu} \cr
&- {i\over \epsilon} {\partial^2 U \over \partial x^\mu\partial x^\nu}
U^{-1} -{i\over \epsilon}{\partial U \over \partial x^\mu}
{\partial U^{-1} \over \partial x^\nu},\cr} \eqno(55)$$
and subtracting the same expression with $\mu, \nu$ reversed. Then add
the result to
$$\eqalign{i\epsilon [{\cal U}'_\mu, {\cal U}'_\nu]&= i\epsilon
  U[{\cal U}_\mu,{\cal U}_\nu]U^{-1}
  + [U{\cal U}_\mu U^{-1}, {\partial U \over \partial x^\nu}U^{-1}] \cr
&+ [{\partial U \over \partial x^\mu}U^{-1} , U{\cal U}_\nu U^{-1}] 
-{i\over \epsilon} [{\partial U \over \partial x^\mu} U^{-1}, 
{\partial U \over \partial x^\nu} U^{-1}] \cr}\eqno(56)$$ 
Whenever the combination
$$ U^{-1} {\partial U \over \partial x^\mu} U^{-1} $$
appears, it should be replaced by
$$ -{\partial U^{-1} \over \partial x^\mu}.$$
The result $(54)$ then follows after a little manipulation.
\par Now, consider the possibility that this finite gauge transformation
leaves ${\cal U}_\mu {\cal U}^\mu = -1$.

\par We write out
$$ \eqalign{(U{\cal U}_\mu U^{-1} - {i\over \epsilon} {\partial U
\over 
\partial  x^\mu} U^{-1})(U{\cal U}^\mu U^{-1} - {i\over \epsilon} 
{\partial U \over \partial
  x_\mu} U^{-1})&= -1 -{i \over \epsilon}{\partial U \over \partial
  x^\mu}{\cal U}^\mu U^{-1}\cr & - {i\over \epsilon} U{\cal U}_\mu
  U^{-1} {\partial U\over
    \partial x_\mu} U^{-1} \cr &- { 1 \over \epsilon^2} {\partial U
    \over \partial x^\mu} U^{-1} {\partial U \over \partial
    x_\mu}U^{-1} \cr
&= -1 -{i\over \epsilon} {\partial U \over \partial x^\mu} {\cal U}^\mu
U^{-1} + {i \over \epsilon} U {\cal U}_\mu {\partial U^{-1} \over \partial
  x_\mu}\cr &+ {1 \over \epsilon^2} {\partial U \over \partial x^\mu}
{\partial U^{-1} \over \partial x_\mu}.\cr} \eqno(57)$$
\par It may be  possible that $U$ can be chosen to make all but the
  first term in $(57)$ vanish, but in the case of finite gauge 
transformations, it is not so easy to see how to construct examples.  For 
the infinitesimal case, it is, however,
easy to construct a gauge function with the required properties.   For
$$ U \cong 1+ iG, \eqno(58)$$
where $G$ is infinitesimal, $(51)$ becomes
$$ {\cal U}'_\mu = {\cal U}_\mu + i [G,{\cal U}_\mu] 
+ {1\over \epsilon}{\partial G \over
  \partial x^\mu} + O(G^2). \eqno(59)$$  
Then,
$$\eqalign{ {\cal U}'_\mu {\cal U}'^\mu &\cong {\cal U}_\mu n_\mu 
+i ({\cal U}_\mu [G,{\cal U}^\mu] +
 [G,{\cal U}_\mu]{\cal U}^\mu)\cr 
&+ { 1\over \epsilon} \bigl( {\partial G \over \partial x^\mu}{\cal U}^\mu + 
{\cal U}_\mu {\partial G \over \partial x_\mu} \bigr).\cr} \eqno(60)$$ 
\par Let us take  
$$ \eqalign{G &= -{i\epsilon \over 2}\sum \bigl\{\omega_{\lambda
\gamma}({\cal U},x), 
({\cal U}^\lambda{\partial \over \partial
  {\cal U}_\gamma} - {\cal U}^\gamma {\partial \over \partial 
{\cal U}_\lambda})\bigr\}\cr 
&\equiv {\epsilon \over 2}\sum \bigl\{\omega_{\lambda
\gamma}({\cal U},x),N^{\lambda\gamma}
\bigr\} \cr} \eqno(61)$$
where symmetrization is required since $\omega_{\lambda \gamma}$ is a 
function of ${\cal U}$ as well as $x$, and 
$$ N^{\lambda\gamma}= -i ({\cal U}^\lambda {\partial \over \partial 
{\cal U}_\gamma}
  - {\cal U}^\gamma {\partial\over \partial {\cal U}^\lambda}). \eqno(62)$$
\par This construction is valid 
in the initially special gauge, which we shall call 
the ``special abelian gauge'', in which the  components of ${\cal
U}^\mu$ commute.  The appearance of ${\cal U}^\mu$ in the gauge functions
 is then admissible
since this quantity acts on the wave functions
$<{\cal U},x|\psi)=\psi({\cal U},x)$ at the point $x$,in the
representation 
in which 
the operator ${\cal U}^\mu$ on $L^2(d{\cal U})$ is diagonal. 
\par Our investigation in the following will be concerned with a study of the 
infinitesimal gauge neighborhood of this limit, where the components of 
${\cal U}^\mu$ do not commute, and therefore constutite a Yang Mills type field.  
We shall show in the limit that the corresponding  field equations acquire
nonlinear terms, and may therefore suppress the caustic singularities
found by Contaldi {\it et al} [10].  They found that nonlinear terms associated with a
 non-Maxwellian type action, such as $ (\partial_\mu {\cal U}^\mu )^2$,
 could 
avoid this caustic singularity, so that the nonlinear terms we find as a residue of the 
Yang-Mills structure induced by our gauge transformation might achieve this 
effect in a natural way. 
\par   The second term of $(60)$, which is the
commutator of $G$ with ${\cal U}^\mu {\cal U}_\mu$ vanishes, since this product is
Lorentz invariant (the symmetrization in $G$ does not affect this result). 
\par We now consider the third term in $(60)$.
$$\eqalign{ {1 \over \epsilon}\bigl( {\partial G \over \partial
 x^\mu}{\cal U}^\mu
 + {\cal U}_\mu {\partial G \over 
\partial x_\mu}\bigr)&= {1\over 2}\bigl\{ {\partial \omega_{\lambda
\gamma}\over \partial x^\mu},N^{\lambda\gamma}\bigr\}{\cal U}^\mu 
+ {\cal U}^mu \bigl\{ {\partial \omega_{\lambda
\gamma}\over \partial x^\mu},N^{\lambda\gamma}\bigr\}\cr &=
{1 \over 2}\bigl\{N^{\lambda\gamma}{\partial \omega_{\lambda
\gamma}\over \partial x^\mu}{\cal U}^\mu  + {\partial \omega_{\lambda
\gamma}\over \partial x^\mu}N^{\lambda \gamma} {\cal U}^\mu\cr
&+{\cal U}^\mu N^{\lambda\gamma}{\partial \omega_{\lambda
\gamma}\over \partial x^\mu}  + {\cal U}^\mu{\partial \omega_{\lambda
\gamma}\over \partial x^\mu}N^{\lambda \gamma}\bigr\} \cr}\eqno(63)$$
There are two terms proportional to 
$${\partial \omega_{\lambda
\gamma}\over \partial x^\mu}{\cal U}^\mu.$$
If we take (locally)
$$ \omega_{\lambda \gamma}({\cal U},x) = \omega_{\lambda \gamma}(k_\nu
x^\nu),\eqno(64)$$ 
where $k_\nu {\cal U}^\nu =0$, then 
$${\partial \omega_{\lambda \gamma} \over \partial x^\mu}{\cal U}_\mu
 = k_\mu
 {\cal U}^\mu
\omega'_{\lambda \gamma}=0.\eqno(65)$$
For the remaining two terms,
$$\eqalign{{\cal U}^\mu N^{\lambda \gamma} {\partial \omega_{\lambda \gamma}
 \over 
\partial x^\mu} &+
 {\partial \omega_{\lambda \gamma} \over \partial x^\mu}N^{\lambda \gamma}
 {\cal U}^\mu \cr &=N^{\lambda \gamma}{\cal U}^\mu {\partial \omega_{\lambda \gamma} \over 
\partial x^\mu}\cr
&+[ {\cal U}^\mu,N^{\lambda \gamma}] {\partial \omega_{\lambda \gamma} \over
\partial x^\mu}+  {\partial \omega_{\lambda \gamma} \over \partial
x^\mu}
{\cal U}^\mu 
N^{\lambda \gamma}\cr &+  {\partial \omega_{\lambda \gamma} \over 
\partial x^\mu} 
  [N^{\lambda \gamma},{\cal U}^\mu].\cr}\eqno(66)$$
Since the commutators contain only terms linear in ${\cal U}_\mu$ and they
 have opposite sign, they cancel. The remaining terms are zero by
 the argument $(65)$.  The condition 
${\cal U}_\mu {\cal U}^\mu = -1$ is therefore
 invariant under this gauge transformation, involving the coefficient
 $\omega_{\lambda \gamma}$ which is a function of the projection of
 $x^\mu$ onto a hyperplane orthogonal to ${\cal U}_\mu$, {\it i.e.}, a
function of $k_\mu x^\mu$, where  $k_\mu {\cal U}^\mu = 0$.  The vector
 $k_\mu$, of course, depends on ${\cal U}_\mu$ (for 
example, $k_\mu ={\cal U}_\mu ({\cal U}\cdot b)+ b_\mu $,
 for some $b_\mu \neq 0$).
 \par We now demonstrate explicitly the nonabelian nature of the gauge fields
after infinitesinal gauge transformation. 
 With $(59)$, the commutator term in $(53)$ is
$$\eqalign{[{\cal U}'_\mu, {\cal U}'_\nu]&= ( {\cal U}_\mu + 
i [G,{\cal U}_\mu] 
+ {1\over \epsilon}{\partial G \over
  \partial x^\mu})( {\cal U}_\nu + i [G,{\cal U}_\nu] + {1\over \epsilon}{\partial G \over
  \partial x^\nu})\cr &-( {\cal U}_\nu + i [G,{\cal U}_\nu] + 
{1\over \epsilon}{\partial G \over
  \partial x^\nu})( {\cal U}_\mu + i [G,{\cal U}_\mu] + {1\over \epsilon}{\partial G \over
  \partial x^\mu})\cr
&= {1\over \epsilon}\bigl\{ [{\cal U}_\mu, {\partial G \over \partial x^\nu}] -
[{\cal U}_\nu, {\partial G \over \partial x^\mu}]\bigr\}\cr
&+i[{\cal U}_\mu,[G,{\cal U}_\nu]]-i[{\cal U}_\nu,[G,{\cal U}_\mu]],\cr} \eqno(67)$$
where the remaining terms have identically cancelled. Note that
this expression does not contain any noncommutative quantities.
 Now,
$$ [G,{\cal U}_\nu] = 2i\epsilon{\omega_\nu}^\gamma {\cal U}_\gamma \eqno(68)$$
and 
$$ [{\cal U}_\mu, {\partial G \over \partial x^\nu}] = 2i\epsilon
{\cal U}_\lambda 
{\partial {\omega^\lambda}_\mu \over \partial x^\nu}. \eqno(69)$$
\par The terms involving $[G,{\cal U}_\nu]$ and $[G,{\cal U}_\mu]$
therefore cancel, so that
$$[{\cal U}'_\mu, {\cal U}'_\nu] = 2i {\cal U}_\lambda
\bigl({\partial {\omega^\lambda}_\mu \over \partial x^\nu}-
{\partial {\omega^\lambda}_\nu \over \partial x^\mu}\bigr)\eqno(70)$$
We have taken ${\omega^\lambda}_\mu ={\omega^\lambda}_\mu(k_\sigma
x^\sigma)$,
so that
 $${\partial {\omega^\lambda}^\mu \over \partial x^\nu}= k_\nu
 {\omega'^\lambda}_\mu , \eqno(71)$$
and therefore
$$[{\cal U}'_\mu, {\cal U}'_\nu] = 2i ( k_\nu
 {\omega'^\lambda}_\mu -  k_\mu {\omega'^\lambda}_\nu){\cal
 U}_\lambda, 
\eqno(72)$$
generally not zero. This demonstrates the nonabelian character of the fields.
In the Abelian limit, we may take $\omega' \rightarrow 0$, but as we
shall a residual nonlinearity, which depends on $\omega''$ may  remain
in the field equations
\par We now consider the derivation of field equations from a
 Lagrangian constructed
with the $\psi$'s and $f^{\mu\nu}f_{\mu\nu}$. We take the Lagrangian
to be of the form (the indices are raised and lowered with
$g_{\mu\nu}$)
$${\cal L} = {\cal L}_f + {\cal L}_m, \eqno(73)$$
where 
$${\cal L}_f = -{1 \over 4} f^{\mu\nu} f_{\mu\nu} \eqno(74)$$
and
$${\cal L}_m = \psi^*\bigl( i {\partial \over \partial
\tau} - {1 \over 2M}
(p_\mu - \epsilon {\cal U}_\mu)g^{\mu \nu}
(p_\nu - \epsilon {\cal U}_\nu) -
\Phi\bigr) \psi \ \ \ \ +\ \ \ {\rm c.c.} \eqno(75)$$
\par We shall be working in the infinitesimal neighborhood of the
special gauge for Abelian ${\cal U}_\mu$, for which it has the form given in
$(59)$ for infinitesinal $G$.  It is therefore not Abelian to first
order, but we take its variation $\delta {\cal U}$ to be a c-number function,
carrying the variation, to lowest order, by variation of the first
term in $(59)$, and not varying the part of ${\cal U}$ introduced by the
infinitesimal gauge transformation (evaluated on the original value of
${\cal U}$).
\par In carrying out the variation of ${\cal L}_m$, the contributions
of varying the $\psi$'s with respect to ${\cal U}$ vanish due to the field
equations (Stueckelberg-Schr\"odinger equation) obtained by varying
$\psi^*$ (or $\psi$), and therefore in the variaton with respect to
${\cal U}$, only the explicit presence of ${\cal U}$ in $(75)$ need be
taken 
into account.
\par Note that for the general case of ${\cal U}$ generally operator valued,
we can write
$$ \psi^* (p_\mu - \epsilon {\cal U}_\mu)g^{\mu \nu}
(p_\nu - \epsilon {\cal U}_\nu)\psi
= g^{\mu \nu}\bigl((p^\mu - \epsilon {\cal U}^\mu)\psi \bigr)^* 
(p_\nu - \epsilon{\cal U}_\nu)\psi, \eqno(76)$$
since the Lagrangian density $(75)$ contains an integration over
$(d{\cal U}')(d{\cal U}'')$ (considered in lowest order) as well as an 
integration
over $(dx)$ in the action and the operators ${\cal U}$ are Hermitian. In
the limit in which ${\cal U}$ is evaluated in the special Abelian gauge
(real valued), and noting that $p_\mu$ is represented by an imaginary
differential operator, we
can write this as
$$ g^{\mu \nu}\psi^* (p_\mu - \epsilon {\cal U}_\mu)(p_\nu - \epsilon {\cal U}_\nu)\psi
=-g^{\mu \nu} (p_\mu + \epsilon {\cal U}_\mu)\psi^* (p_\nu - \epsilon
{\cal U}_\nu)\psi, \eqno(77)$$
i.e., replacing explicitly $p_\mu $ by $-i(\partial/\partial x^\mu)\equiv
 -i\partial_\mu $,
we have
$$\delta_{\cal U} {\cal L}_m = -i{\epsilon \over 2M}\bigl\{\psi^* (\partial_\mu -
i\epsilon {\cal U}_\mu)\psi 
- ((\partial_\mu +i \epsilon {\cal U}_\mu)\psi^*) \psi \bigr\}\delta
{\cal U}^\mu,
\eqno(78)$$
where we have called $g^{\mu \nu}\delta {\cal U}_\nu = \delta {\cal U}_\mu$,
 or,
$$\delta_{\cal U} {\cal L}_m = j_\mu ({\cal U},x) \delta {\cal U}^\mu,
\eqno(79)$$
where  $j_\mu$ has the usual form of a gauge invariant current.

\par For the calculation of the variation of ${\cal L}_f$ we note that
the commutator term in $(52)$ is, in lowest order, a c-number function,
as given in $(72)$.
\par Calling 
             $$ {\omega'^\lambda}_\mu {\cal U}_\lambda \equiv v_\mu,  \eqno(80)$$
we compute the variation of 
$$[{\cal U}'_\mu, {\cal U}'_\nu] = 2i ( k_\nu v_\mu -k_\mu v_\nu) \eqno(81)$$
Then, for
$$ \delta_{\cal U} [{\cal U}'_\mu, {\cal U}'_\nu] = \delta_{{\cal
 U}_\gamma}
 {\partial \over
\partial{\cal U}_\gamma}[{\cal U}'_\mu, {\cal U}'_\nu], \eqno(82)$$
we compute
$$ {\partial \over
\partial{\cal U}_\gamma}[{\cal U}'_\mu, {\cal U}'_\nu] = 
2i({\partial k_\nu \over \partial {\cal U}_\gamma} v_\mu + k_\nu {\partial
v_\mu \over \partial {\cal U}_\gamma}) - (\mu \leftrightarrow
\nu)). \eqno(83)$$
With our choice of $k_\nu= {\cal U}_\nu ({\cal U}\cdot b) + b_\nu$,
$${\partial k_\nu \over \partial {\cal U}_\gamma} = 
{\delta_\nu}^\gamma({\cal U}\cdot b) + {\cal U}_\nu b^\gamma, \eqno(84)$$
so that  
$$ \eqalign{{\partial \over\partial {\cal U}_\gamma} [{\cal U}'_\mu,
{\cal U}'_\nu] &=
2i({\delta_\nu}^\gamma({\cal U}\cdot b) + {\cal U}_\nu b_\gamma) v^\mu \cr
&+ k_\nu {\partial
v_\mu \over \partial {\cal U}_\gamma} - (\mu \leftrightarrow \nu))\cr
&\equiv {{\cal O}^\gamma}_{\mu\nu},\cr} \eqno(85)$$
{\it i.e.}
$$ \delta_{\cal U}[{\cal U}'_\mu, {\cal U}'_\nu] = 
{{\cal O}^\gamma}_{\mu\nu}\delta {\cal U}_\gamma \eqno(86)$$
The quantity $v_\mu$ is proportional to the derivative of 
$\omega^\lambda_\mu$. 
In the limit that $\omega,\omega' \rightarrow 0$ (cf. $(81)$),  the second
derivative, $\omega''$ which appears in ${{\cal O}^\gamma}_{\mu\nu} $
may not vanish
(somewhat analogous to the case in gravitional theory when the
connection form vanishes but the curvature does not), so that this
term can contribute in limit to the special Abelian gauge.
\par Returning to the variation of ${\cal L}_f$ in $(74)$, we see that
$$ \delta {\cal L}_f = -\partial^\nu f_{\mu\nu} \delta {\cal U}^\mu + 
2i f_{\mu\nu} \delta
[{\cal U}_\mu, {\cal U}_\nu],\eqno(87)$$ 
 where we have taken into account the fact that
    $ [{\cal U}_\mu, {\cal U}_\nu]$ is a
commuting function, and integrated by parts the derivatives of $\delta
{\cal U}$. With $(86)$ we obtain
$$\delta {\cal L}_f = -\partial^\nu f_{\mu\nu} \delta {\cal U}^\mu +
2i\epsilon f_{\lambda \sigma} {{\cal O}^{\lambda\sigma}}_\mu \delta
{\cal U}^\mu \eqno(88)$$
Since the coefficient of $\delta {\cal U}^\mu$ must vanish, we obtain, with
$(79)$, the Yang-Mills equations for the fields given the source currents
$$ \partial^\nu f_{\mu\nu} = j_\mu -2i\epsilon f_{\lambda\sigma} {{\cal
O}^{\lambda \sigma}}_\mu, \eqno(89)$$
which is nonlinear in the fields ${\cal U}_\mu$, as we have seen, even in the
Abelian limit, where, from $(78)$ and $(79)$, 
$$ j_\mu = -i{\epsilon \over 2M}\bigl\{\psi^* (\partial_\mu -
i\epsilon {\cal U}_\mu)\psi 
- ((\partial_\mu +i \epsilon {\cal U}_\mu)\psi^*) \psi\bigr\}. \eqno(90)$$
\par We point out that this current corresponds to a flow of the
matter field; the absolute square of the wave functions corresponds to
an event density.  The coupling $\epsilon$ is not necessarily charge,
and the fields ${\cal U}$ are not necessarily electromagnetic even in
the Abelian limit. However, the Hamiltonian $(36)$ leads directly to a
Lorentz type force, similar in form to that generated by the 
Hilbert-Einstein
action. The dynamics of this system will be investigated
in a forthcoming paper. 
  
\bigskip
\noindent{\bf 5. Conclusions} 
\par A map of the type discussed in ref.[8], of a Hamiltonian
containing an Einstein metric, generating the connection form of
general relativity, and a world scalar field, representing a
distribution of energy on the spacetime manifold, into a corresponding
Hamiltonian with a conformal metric (and compatible connection form), 
can account for the structure of
the RAQUAL theory of Bekenstein and Milgrom[5]. Furthermore, applying this
correspondence to a Hamiltonian with gauge-type structure, we have
shown that one obtains a non-compact Kaluza-Klein effective metric
which can account for the $TeVeS$ structure of Bekenstein, Sanders
and Milgrom[7].
\par In order to maintain the constraint condition ${\cal U}_\mu {\cal
U}^\mu = -1 $ for the Bekenstein-Sanders fields, under local gauge
 transformations, we have introduced a class of gauge of gauge
 transformations on the underlying quantum theory which acts on the
 Hilbert bundle, quite analogous to that arising in the second
 quantization of the electromagnetic field (where the vector
 potentials and their time derivatives are considered as quantum
 oscillator variables) associated with the values of the gauge
 fields. The action of this class of gauges induces a nonabelian
 structure on the fields, which therefore satisfy Yang-Mills type
 field equations with source currents associated with matter flow. In
 the Abelian limit, these equations contain residual non-linear
 terms which may avoid the caustic singularities found by Contaldi
 {\it et al} for an electromagnetic type gauge field. 
 \par The phenomenological constraints placed on the $TeVeS$
variables in its astrophysical applications and on its MOND limit[16] 
would, in principle, place constraints on the vector and scalar fields
appearing in the corresponding Hamiltonian model, for which the
 additive world scalar field corresponds to an energy distribution not
 associated with electromagnetic radiation.

\bigskip
\noindent {\it Acknowledgements}
\par One of us (L.H.) would like to thank J.D. Bekenstein for a
discussion at the outset of this work, and for helpful remarks, and we
wish to thank S. Shnider, M. Berry, 
E. Calderon. A. Yahalom, J. Levitan, and M. Lewkowicz 
for discussions of the differential geometry and analytical
mechanics underlying the conformal correspondences we have utilized here.

\bigskip
\noindent {\bf References}
\frenchspacing
\item{1.} C.W. Misner, K.S. Thorne and J.A. Wheeler, {\it
Gravitation}, Freeman, San Francisco (1973).
\item{2.} E.C.G. Stueckelberg, Helv. Phys. Acta {\bf 14}, 372, 588
(1941); {\bf 15}, 23 (1942); see also L.P. Horwitz and C. Piron,
Helv. Phys. Acta {\bf 46}, 316 (1973), R.P. Feynman, Phys. Rev. {\bf
80}, 4401 (1950), and J.S. Schwinger, Phys. Rev. {\bf 82}, 664 (1951).
\item{3.} O. Oron and L.P. Horwitz, Found. Phys. {\bf 31}, 951 (2001),
D. Saad, L.P. Horwitz and R.I. Arshansky, Found. of Phys. {\bf 19},
1125 (1989); see also N. Shnerb and L.P. Horwitz, Phys. Rev. A {\bf
48}, 4068 (1993).
\item{4.} L.P. Horwitz, J. Levitan, M. Lewkowicz, M. Shiffer and
Y. Ben Zion, Phys. Rev. Lett. {\bf 98}, 234301 (2007).
\item{5.}  J.D. Bekenstein and M. Milgrom, Astrophys. Jour. {\bf
286},7 (1984); J.D. Bekenstein in {\it Second Canadian Conference on
General Relativity and Relativistic Astrophysics,}, A. Coley, C. Dyer
and T. Tupper, eds., World Scientific, Singapore (1992).
\item{6.} M. Milgrom, Astrophys. Jour. {\bf 270}, 365, 371, 384
(1983); Ann. of
Phys. {\bf 229}, 384 (1994); Astrophys. Jour. {\bf 287},
571 (1984); Astrophys. Jour. {\bf 302}, 617 (1986).
\item{7.} R.H. Sanders, Astroph. Jour. {\bf 480}, 492 (1997);
J.D. Bekenstein, Phys. Rev. D{\bf 70}, 083509 (2004); J.D. Bekenstein, 
{\it Modified Gravity vs. Dark
Matter: Relativistic Theory for MOND}, 28th Johns Hopkins Workshop on
Current Problems in Particle Theory, June 5-8, 2004,
arXiv:astro-ph/0412652 (2005).
\item{8.} A. Gershon and L.P. Horwitz, {\it Kaluza-Klein Theory as a 
Dynamics in a Dual Geometry} Jour. Math. Phys. (in print).
\item{9.} C.N. Yang and R.L. Mills, Phys. Rev. {\bf 96}, 191 (1954).
\item{10.}  C.R. Contaldi, T. Wiseman and B. Withers, arXiv:0802.1215, 
Phys. Rev. D78(2008) 044034.
\item{11.} L.P. Horwitz, J. Levitan, A. Yahalom and M. Lewkowicz, {\it
Variational calculus for classical dynamics on dual manifolds}, in
preparation.
\item{12.} E. Calderon, R. Kupferman, S. Shnider and L.P. Horwitz, in preparation;
E. Calderon, M.Sc. Thesis, {\it Geometric Formulation of Classical Dynamics and
Hamiltonian Chaos}, Hebrew University, Jerusalem, June 1, 2009.
\item{13.} P.S. Wesson, {\it Space-Time-Matter}, World Scientific,
Singapore (2007);
 {\it Five Dimensional Physics}, World Scientific,
Singapore (2006).
\item{14.} T. Liko, Phys. Lett. B {\bf 617}, 193 (2005). See also
N.E. Mavromatos and M. Sakellariadou, Phys. Lett. B {\bf 652}, 97
(2007) for a study of the possibility of deriving the  {\it TeVeS}
theory from string theory. 
\item{15.} T. Kaluza, Sitz. Oreuss. Akad. Wiss. {\bf 33}, 966 (1921).
\item{16.} E Sagi, Phys. Rev. D {\bf 80}, 044032 (2009) and 
arXiv:1001.155 [gr-qc] 11 Jan. 2010.

\vfill
\end